\documentclass[amssymb]{revtex4}

\usepackage{amsmath,amssymb,setspace}
\usepackage{graphicx}
\usepackage{bm, bbm}      
\usepackage{hyperref}
\hypersetup{
    colorlinks=true,       
    linkcolor=blue,          
    citecolor=blue,        
    filecolor=magenta,      
    urlcolor=cyan,           
    runcolor=cyan
}
\begin{document}

\title{Exponential Enhancement of the Efficiency of Quantum Annealing\\
by Non-Stoquastic Hamiltonians}

\author{Hidetoshi Nishimori and Kabuki Takada}
\affiliation{Department of Physics, Tokyo Institute of Technology, Oh-okayama, Meguro-ku, Tokyo 152-8551, Japan}
\date{\today}

\begin{abstract}
Non-stoquastic Hamiltonians have both positive and negative signs in off-diagonal elements in their matrix representation in the standard computational basis and thus cannot be simulated efficiently by the standard quantum Monte Carlo method due to the sign problem.  We describe our analytical studies of this type of Hamiltonians with infinite-range non-random as well as random interactions from the perspective of possible enhancement of the efficiency of quantum annealing or adiabatic quantum computing.  It is shown that multi-body transverse interactions like $XX$ and $XXXXX$ with positive coefficients appended to a stoquastic transverse-field Ising model render the Hamiltonian non-stoquastic and reduce a first-order quantum phase transition in the simple transverse-field case to a second-order transition.  This implies that the efficiency of quantum annealing is exponentially enhanced, because a first-order transition has an exponentially small energy gap (and therefore exponentially long computation time) whereas a second-order transition has a polynomially decaying gap (polynomial computation time).  The examples presented here represent rare instances where strong quantum effects, in the sense that they cannot be efficiently simulated in the standard quantum Monte Carlo, have analytically been shown to exponentially enhance the efficiency of quantum annealing for combinatorial optimization problems. 
\end{abstract}

\maketitle

\section{Introduction}

Quantum annealing is a metaheuristic for combinatorial optimization problems \cite{Kadowaki1998,Kadowaki1998b,Brooke1999,Farhi2000,Farhi2001,Santoro2002,Santoro2006,Das2008,Morita2008}.  A combinatorial optimization problem can generally be expressed as the minimization of an Ising Hamiltonian, {\em i.e.}, the ground-state search of a classical Ising model \cite{Lucas2014}. Then, quantum fluctuations are appended, typically as a uniform transverse field, and the total Hamiltonian constitutes the transverse-field Ising model. The amplitude of the appended term for quantum fluctuations is gradually decreased from a very large value, large relative to the original classical Ising model, toward zero. If one starts from the ground state of the initial Hamiltonian and the rate of change of the amplitude is sufficiently slow, the system follows the instantaneous ground state according to the adiabatic theorem of quantum mechanics. This implies that the system eventually reaches the ground state of the original Ising model representing the solution to the combinatorial optimization problem. There exists a large body of analytical, numerical, and experimental studies on quantum annealing, and active debates are going on to compare quantum annealing with the corresponding classical heuristic, simulated annealing, recent examples of which include Refs.~ \cite{Boixo2014,Ronnow2014,Katzgraber2014,Hen2015,Heim2015,Isakov2015,Venturelli2015,Katzgraber2015,Steiger2015,Albash2015,Martin-mayor2015,Muthukrishnan2016,Kechedzhi2016,Denchev2016,Mandra2016,Marshall2016,Matsuda2009,Mandra2016b,Crosson2016,Young2010,Hen2011,Farhi2012}.

To numerically test the performance of quantum annealing, one often uses quantum Monte Carlo simulation, which is a classical algorithm to sample the equilibrium distribution of the transverse-field Ising model. Although the quantum Monte Carlo simulation is designed to sample the equilibrium Boltzmann distribution, it has been found that some aspects of dynamics of quantum annealing can also be described by quantum Monte Carlo simulations \cite{Isakov2015,Jiang2016,Denchev2016}.  Also remarkable are the generic convergence conditions for quantum annealing under quantum dynamics \cite{Morita2007,Somma2007,Morita2008} and quantum Monte Carlo simulations \cite{Morita2006,Morita2008}, both of which have a very similar asymptotic polynomial decrease of the control parameter that is much quicker than the corresponding inverse-log law for simulated annealing \cite{Geman1984}. These observations suggest the possibility that quantum annealing might be efficiently simulated on classical computers even for its dynamical aspects, the latter being important to judge the performance of quantum annealing.  If this is indeed the case, the role of dedicated hardware to run quantum annealing may have to be reconsidered.

Related to the above observation is the concept of stoquastic Hamiltonians \cite{Bravyi2008}.  Loosely speaking, it is a class of Hamiltonians that can usually be simulated efficiently on classical computers because there is no sign problem in the standard classical implementation using the Suzuki-Trotter decomposition \cite{Suzuki1976} \footnote{It is to be noticed that, in some cases, it is non-trivial to efficiently simulate a stoquastic Hamiltonian. See, for example, \cite{Hastings2013,Jarret2016}.}.   More formally, a stoquastic Hamiltonian has off-diagonal elements all non-positive in the standard computational basis to diagonalize the $z$ component of the Pauli matrix at each site $i$. The transverse-field Ising model belongs to this category.  A non-stoquastic Hamiltonian, by contrast, has both signs in the off-diagonal elements, which causes negative signs in the effective Boltzmann factors when Trotter-decomposed to run simulations on classical computers. This means that it is practically impossible to classically simulate non-stoquastic Hamiltonians by the standard method.  It may then be the case that a proper term added to a stoquastic Hamiltonian, which causes both signs in the matrix representation in the computational basis, represents strong quantum effects not to be classically simulated in a straightforward manner.  Such a term might lead to enhanced performance of quantum annealing as compared to the conventional method with the stoquastic transverse-field Ising model. In this relation, it is to be noticed that the transverse-field Ising model with longitudinal fields can be universal in quantum computation if $XX$ interactions are added with appropriate coefficients \cite{Biamonte2008}.

There exist several studies related to this idea. Farhi {\em et al.} \cite{Farhi2002} investigated the effects of randomly generated non-stoquastic Hamiltonians in a variant of the infinite-range Ising model and found that a finite fraction of examples showed enhancement of performance compared to the stoquastic case.  Crosson {\em et al.} \cite{Crosson2014} ran extensive numerical tests of hard MAX-2SAT problems by directly solving the Schr\"odinger equation for small-size systems.  They concluded that additional terms, which make the Hamiltonian non-stoquastic, improve the success rate, although not decisively better than stoquastic cases.  Hormozi {\em et al.} \cite{Hormozi2016} numerically studied the spin glass problem to find that non-stoquastic Hamiltonians have improved success probabilities for hard instances, possibly not by increasing the energy gap for strict adiabatic evolution but by promoting diabatic transitions.  Seki and Nishimori \cite{Seki2012,Seki2015} and Seoane and Nishimori \cite{Seoane2012} used quantum statistical-mechanical techniques to analyze systematically the infinite-range Ising models with ferromagnetic as well as random interactions to conclude that additional terms, by which the Hamiltonian becomes non-stoquastic, sometimes reduce first-order quantum phase transitions in the stoquastic Hamiltonian to second-order transitions. This means an exponential enhancement of the efficiency, exponential in the system size, because second-order quantum phase transitions have the minimum energy gap that decreases polynomially as a function of the system size whereas first-order transitions have an exponentially small gap.  Remember that the adiabatic theorem states that the time needed for a system to stay close to the instantaneous ground state is proportional to the inverse of a polynomial of the minimum energy gap \cite{Jansen2007,Lidar2009,Elgart2012}.

The present article describes the findings in Refs. \cite{Seki2012,Seki2015,Seoane2012} from the viewpoint of possible enhancement of the efficiency by non-stoquastic Hamiltonians, which was not mentioned explicitly in those papers.  Also explained is the effect of interactions of the system with its environment.

\section{Ferromagnetic $p$-spin model}

The first problem to be discussed is the ferromagnetic $p$-spin model with infinite-range interactions \cite{Seki2012,Jorg2010},
\begin{equation}
    H_0=-N\left(\frac{1}{N}\sum_{i=1}^N \sigma_i^z\right)^p,
    \label{p-spin-Hamiltonian}
\end{equation}
where $\sigma_i^z$ is the $z$ component of the Pauli matrix at site $i(=1,2,\cdots, N)$, and $p(\ge 3)$ is an integer.  The ground state of this Ising Hamiltonian is doubly degenerate for $p$ even, $\sigma_i^z=1~(\forall i)$ and $\sigma_i^z=-1~(\forall i)$, and non-degenerate for $p$ odd, $\sigma_i^z=1~(\forall i)$.  This Hamiltonian (the cost function for combinatorial optimization) is a simple polynomial of the order parameter $m=\big(\sum_{i=1}^N\sigma_i^z\big)/N$, and the steepest descent method readily finds the ground state.  In this sense, the problem is easily solved classically.  Our focus, however, is on how quantum annealing compares with its classical counterpart, simulated annealing, according to the criterion of `limited quantum speedup' \cite{Ronnow2014} as well as on how a non-stoquastic Hamiltonian compares with its stoquastic counterpart.

In the present section, we consider the case with $p\ge 3$ since the $p=2$ model in a transverse field has a second-order quantum phase transition and is therefore easy to solve already in the stoquastic case.  We restrict ourselves to the subspace of $m\ge 0$ without losing generality.

\subsection{Simulated annealing}

\begin{figure}[h!]
\begin{center}
\includegraphics[width=0.4 \linewidth]{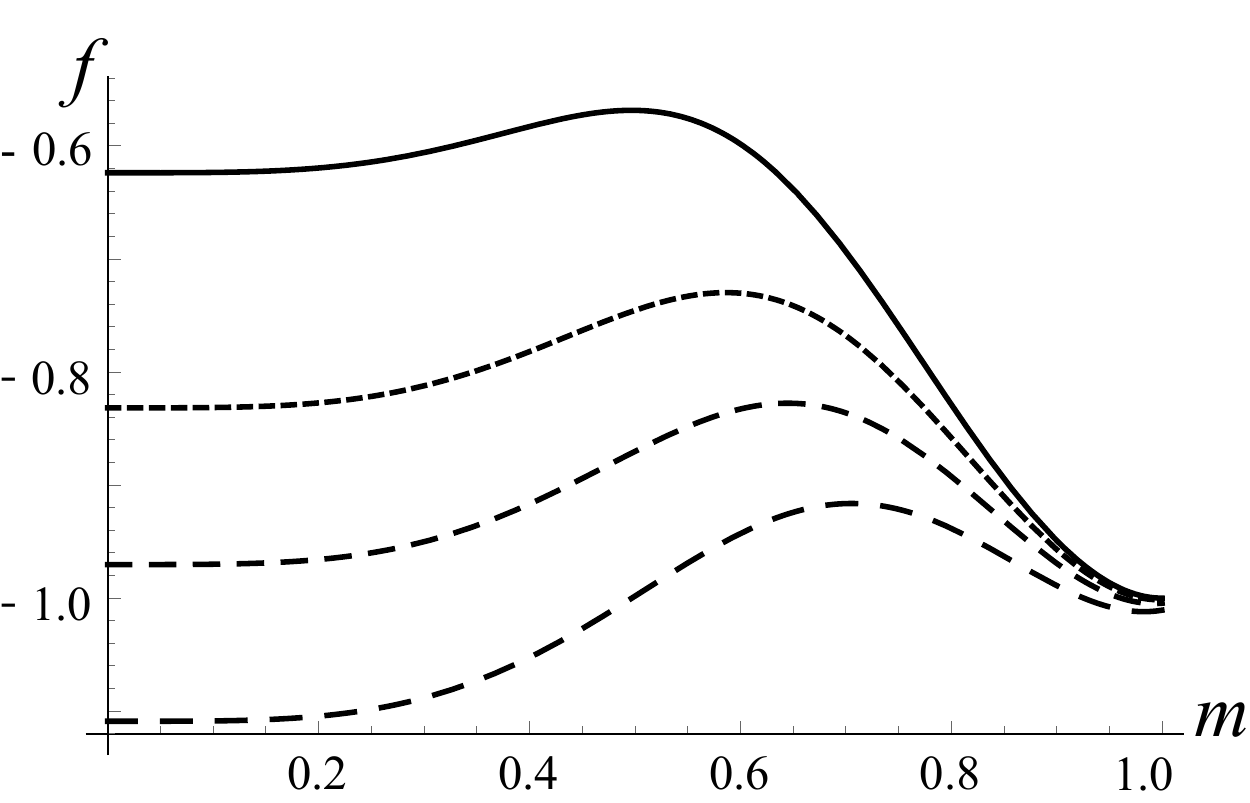}
\end{center}
\caption{Free energy per site $f(m)$ as a function of the order parameter $m$ for $p=4$ at four values of the temperature, $T=0.9, 1.5, 1.8$ and 2.2, from top to bottom. The minimum jumps between $m=0$ to $m\approx 1$ as the temperature changes.}\label{fig:SA_free_energy_p=4}
\end{figure}

Simulated annealing is a classical heuristic to sample the Boltzmann distribution with the temperature decreasing from a very high value to zero \cite{Kirkpatrick1983}. To understand the process theoretically for the present problem, it is convenient to see how the free-energy landscape behaves as a function of the order parameter at each given temperature.  To this end, we write the partition function as
\begin{align}
    Z&={\rm Tr}\int dm\, \delta\big(Nm-\sum_i\sigma_i^z\big) e^{\beta Nm^p}\nonumber\\
    &={\rm Tr}\int dm\,d\tilde{m}\, \exp\Big(-i\tilde{m}\big(Nm-\sum_i \sigma_i^z\big)+\beta Nm^p\Big)\nonumber\\
    &=\int dm\, d\tilde{m}\, \exp\big(-iN\tilde{m}m+\beta Nm^p+N\ln 2\cosh (i\tilde{m})\big),
\end{align}
where $\beta=1/T$ is the inverse temperature with the Boltzmann constant chosen to be 1 for simplicity. We have dropped a trivial prefactor $1/2\pi$ in the above expression.  The exponent in the integrand in the last line is the generalized free energy $-\beta Nf(m,\tilde{m})$ for given values of $m$ and $\tilde{m}$.  In the thermodynamic limit $N\to\infty$, the integral is evaluated by the saddle point method.  The extremum condition of the exponent with respect to $m$ is $-i\tilde{m}+\beta pm^{p-1}=0$. By eliminating $\tilde{m}$ using this equation, we obtain the Landau-type free energy per site as a function of $m$,
\begin{equation}
f(m)=(p-1)m^p-T \ln 2\cosh \big(\beta pm^{p-1}\big).
\end{equation}
As shown in Fig. \ref{fig:SA_free_energy_p=4}, there exists a jump in the minimum as the temperature changes and hence the transition is of first order.
In simulated annealing, one should drive the system from a disordered state ($m=0$) to an ordered state $m>0$ over the free-energy barrier as the temperature is decreased. This takes an exponentially long time since the probability to go over the peak of the free-energy barrier is exponentially small, proportional to $\exp ( -N\beta \Delta f)$, where $\Delta f$ is the height of the barrier of the free energy per site at the transition temperature.  Therefore, the present simple problem is hard to solve by simulated annealing. The existence of a first-order phase transition is the origin of the difficulty.

\subsection{Quantum annealing with stoquastic Hamiltonian}

What will happen if we apply quantum annealing to the same problem? The conventional choice of the transverse-field Ising model for quantum annealing has the Hamiltonian
\begin{equation}
H(s)=s H_0\big(\{\sigma_i^z\}\big)-(1-s)\sum_{i=1}^N \sigma_i^x,
\label{stoquastic_H_generic}
\end{equation}
where $s$ is the time-dependent parameter to control the dynamical evolution of the system running from the initial value $s(t=0)=0$ to the final $s(t=\tau)=1$ with $\tau$ being the computation time. A typical example is $s=t/\tau$. Equation (\ref{stoquastic_H_generic}) is a stoquastic Hamiltonian.

For our problem Hamiltonian of Eq. (\ref{p-spin-Hamiltonian}), Eq. (\ref{stoquastic_H_generic}) reads
\begin{equation}
H(s)=-s N\left(\frac{1}{N}\sum_{i=1}^N \sigma_i^z\right)^p-(1-s)\sum_{i=1}^N \sigma_i^x,
\label{TFIM}
\end{equation}
which can be expressed in terms of the normalized total spin operators,
\begin{equation}
m_z=\frac{1}{N}\sum_{i=1}^N\sigma_i^z,\quad m_x=\frac{1}{N}\sum_{i=1}^N\sigma_i^x
\end{equation}
as
\begin{equation}
 H=-sN(m_z)^p-(1-s)N m_x.
\end{equation}
The normalized total spin operators satisfy the commutation relation,
\begin{equation}
\big[m_x,m_z\big]=-\frac{2i}{N}\,m_y.
\label{commutation}
\end{equation}
Since the norm of those operators, defined as the largest absolute eigenvalue, is unity, the right-hand side of Eq. (\ref{commutation}) vanishes in the thermodynamic limit $N\to\infty$. The same is true for other commutators of $m_x, m_y$, and $m_z$.  It is also useful to remember that the total spin operator commutes with the Hamiltonian and therefore is conserved in the present $p$-spin model.  We are interested in the subspace with the largest value of the total spin, since we start quantum annealing in this subspace. For these reasons, we may regard the operators $m_x, m_y$ and $m_z$ as $x, y$ and $z$ components of a classical vector ${\bm m}$ of unit length written as
\begin{equation}
m_x=\cos \theta,~m_z=\sin\theta \cos \phi
\end{equation}
in the polar coordinate. Equation (\ref{TFIM}) then reduces to a classical energy, whose value per site is
\begin{equation}
e=-s\, \sin^p\theta\cos^p\phi-(1-s)\cos\theta.
\label{classical-e-stoq}
\end{equation}
To minimize this energy, the angle $\phi$ is 0 if $p$ is odd, and $0$ or $\pi$ for $p$ even. We may thus drop $\cos\phi$ and write
\begin{equation}
e=-s\, \sin^p\theta-(1-s)\cos\theta.
\label{classical-e-stoq2}
\end{equation}

\begin{figure}[h!]
\begin{center}
\includegraphics[width=0.4 \linewidth]{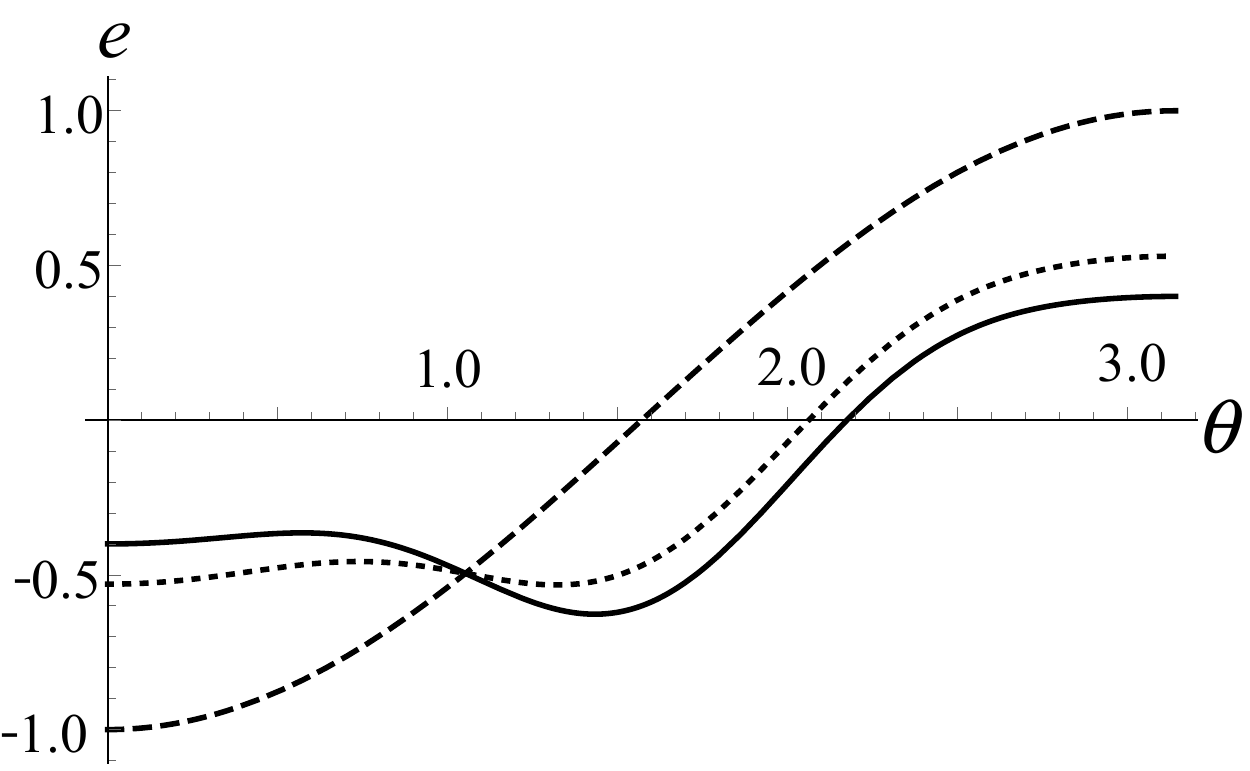}
\end{center}
\caption{Energy per site as a function of the polar angle $\theta$ for $p=5$ at $s=0.2$ (dashed), $s=0.47$ (dotted) and $s=0.6$ (full line). The minimum jumps between $\theta=0$ and  $\theta >0$.}\label{fig:QA_stoquastic_p=5}
\end{figure}

As one sees in Fig. \ref{fig:QA_stoquastic_p=5} for the case of $p=5$, the minimum jumps from $\theta=0$ to $\theta>0$ at some $s$.
The system has a first-order phase transition, and the energy gap between the ground state and the first excited state is expected to decrease exponentially as a function of the system size, which has indeed been shown to be the case explicitly in the present problem both analytically and numerically \cite{Jorg2010}. This is a difficult situation for quantum annealing in its adiabatic formulation (adiabatic quantum computation \cite{Farhi2000,Farhi2001}) because one should spend an exponentially long computation time $\tau$ to reach the correct ground state of the target Hamiltonian $H_0$. The authors of \cite{Jorg2010} thus wrote legitimately that this is ``a problem that quantum annealing cannot solve".

One may wonder if the above analysis using a classical vector would properly describe the essential features of quantum annealing under the Hamiltonian Eq. (\ref{TFIM}).  The answer is positive as far as the properties of phase transitions are concerned:  J{\"{o}}rg {\em et al} used full quantum statistical-mechanical tools to reach the same conclusion as above \cite{Jorg2010}. Quantum effects should be carefully taken into account if one wishes to fully understand the behavior of the energy gap for finite-size systems, as was done by J{\"{o}}rg {\em et al} \cite{Jorg2010}, and to describe more subtle properties of the system around the phase transition and within the ferromagnetic phase \cite{Susa2016}. However, the classical analysis is sufficient to predict the type of phase transitions in the thermodynamic limit.  We take advantage of this observation in the next section for a non-stoquastic Hamiltonian.

It is also worth noticing that the performance of quantum annealing is comparable to that of simulated annealing discussed in the previous section, both of which should spend an exponentially long time to reach the ground state. In this sense, there is no `limited quantum speedup' in the present case, according to the classification of \cite{Ronnow2014}, although there may exist quantitative differences such as the difference in the coefficients of the exponent.

\subsection{Quantum annealing with non-stoquastic Hamiltonian}

We next study the non-stoquastic case with the Hamiltonian
\begin{equation}
 H(s,\lambda)=-s\lambda N \Big(\frac{1}{N}\sum_{i=1}^N\sigma_i^z\Big)^p
 +s(1-\lambda)N\Big(\frac{1}{N}\sum_{i=1}^N\sigma_i^x\Big)^k-(1-s)\sum_{i=1}^N \sigma_i^x,
\end{equation}
where $\lambda \in [0,1]$ is a parameter to control the strength of the additional term, the second term on the right-hand side (to be called the antiferromagnetic multiple-$X$ term), and $k(\ge 2)$ is an integer. The parameter $\lambda$ will later be chosen to be a function of $s$.  Notice that the coefficient of the second term $s(1-\lambda)$ is positive so that this term makes the Hamiltonian non-stoquastic.  For $\lambda =1$, the above Hamiltonian reduces to the stoquastic Eq. (\ref{TFIM})

Quantum annealing starts at $s=0$ ($\lambda$ arbitrary), in which case the Hamiltonian is the simple transverse field,
\begin{equation}
H(0,\lambda )=-\sum_{i=1}^N \sigma_i^x
\end{equation}
just as in the stoquastic case of the previous section. Then one increases $s$ toward 1 and, at the same time, $\lambda$ is increased toward 1 in an appropriate way as will be described later.  The goal is at $s=\lambda =1$, where the final Hamiltonian is the target cost function,
\begin{equation}
 H(1,1)=-N \Big(\frac{1}{N}\sum_{i=1}^N\sigma_i^z\Big)^p.
\end{equation}

The analysis proceeds as before by the replacement of the normalized total spin operator with a classical unit vector.  The energy per site is
\begin{equation}
 e=-s\lambda \sin^p\theta +s(1-\lambda)\cos^k \theta -(1-s)\cos\theta.
 \label{energy_nonstoquastic}
\end{equation}

\begin{figure}[h!]
\begin{center}
\includegraphics[width=0.4 \linewidth]{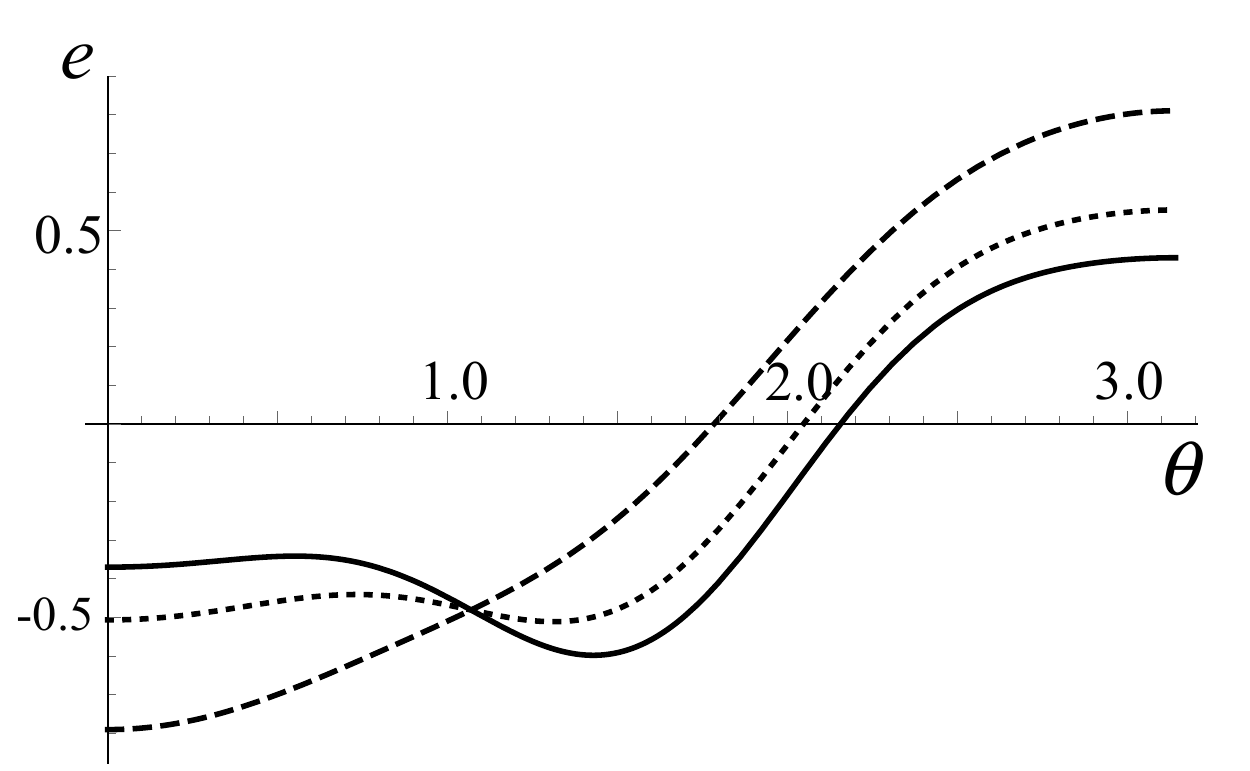}
\end{center}
\caption{Energy per site for the non-stoquastic Hamiltonian with $p=5, k=2$ and $\lambda =0.95$ at $s=0.2$ (dashed), 0.47 (dotted) and 0.6 (full line). A first-order phase transition happens at $s=0.47$}\label{fig:QA_nonstoquastic_p=5a}
\end{figure}

\begin{figure}[h!]
\begin{center}
\includegraphics[width=0.4 \linewidth]{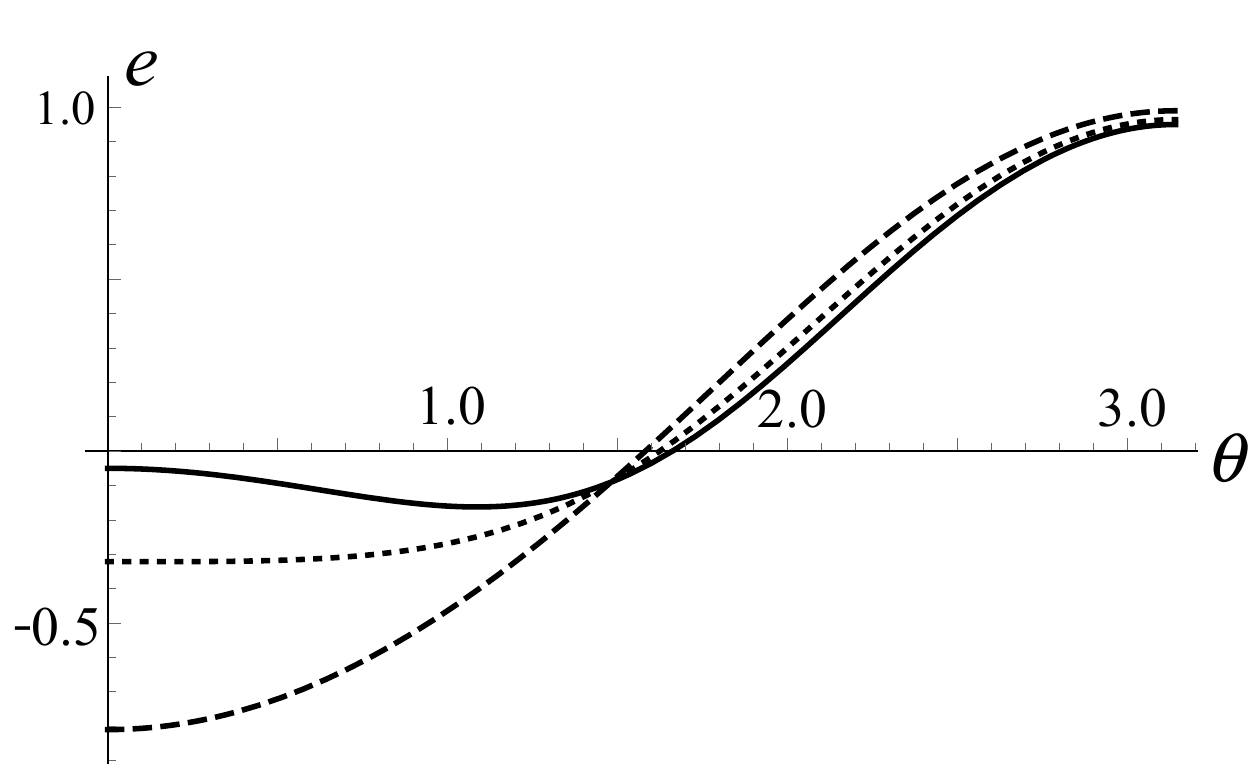}
\end{center}
\caption{Energy per site for the non-stoquastic Hamiltonian with $p=5, k=2$ and $\lambda =0.1$ at $s=0.1$ (dashed), 0.357 (dotted) and 0.5 (full line). The phase transition at $s=0.357$ is of second order.}\label{fig:QA_nonstoquastic_p=5b}
\end{figure}

Two typical examples of the behavior of this energy at $p=5$ and $k=2$ are shown in Fig. \ref{fig:QA_nonstoquastic_p=5a} for $\lambda=0.95$ and Fig. \ref{fig:QA_nonstoquastic_p=5b} for $\lambda=0.1$.  The former is essentially the same as the stoquastic case ($\lambda =1$) of Fig. \ref{fig:QA_stoquastic_p=5} with a first-order phase transition at $s=0.47$.  The latter is drastically different with a second-order phase transition at $s=0.357$.
This second-order transition point can be understood by a Landau-type expansion of the energy near $\theta=0$,
\begin{equation}
e\approx -(1-2s+s\lambda)+\frac{1-3s+2s\lambda}{2}\, \theta^2.
\end{equation}
A second-order phase transition takes place when the coefficient of the quadratic term vanishes according to the Landau theory \cite{Nishimori_book}, $s=1/(3-2\lambda)$, which gives $s=0.357$ for $\lambda=0.1$. This second-order transition is masked by a first-order transition if the latter happens at a smaller $s$, which is indeed the case for $\lambda=0.95$

\begin{figure}[h!]
\begin{center}
\includegraphics[width=.35\textwidth]{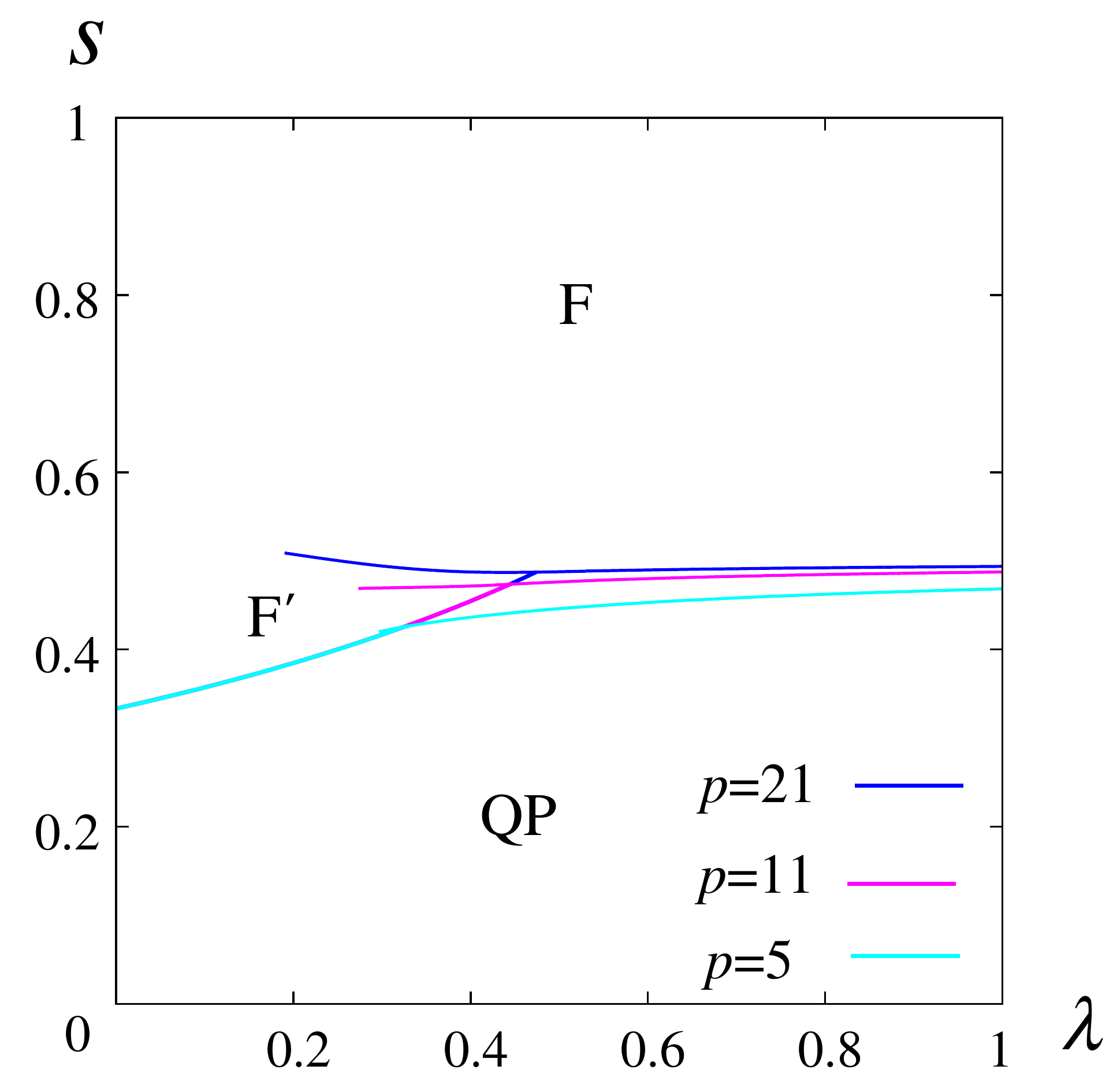}
\end{center}
\caption{$\lambda$-$s$ phase diagram of the non-stoquastic Hamiltonian with $k=2$ (antiferromagnetic $XX$ interactions). The line separating QP (quantum paramagnetic) and F$'$ (ferromagnetic) phases represents second order phase transitions, and all other lines are for first-order transitions. There is a line of first-order phase transitions within the ferromagnetic phase, and thus labels F and F$'$ are given to distinguish the two ferromagnetic phases although they have no qualitative difference. The axis $\lambda=1$ on the right of the panel corresponds to the stoquastic case.}\label{fig:nonstoquastic_phase_diagram_k2}
\end{figure}

\begin{figure}[h!]
\begin{center}
\includegraphics[width=.35\textwidth]{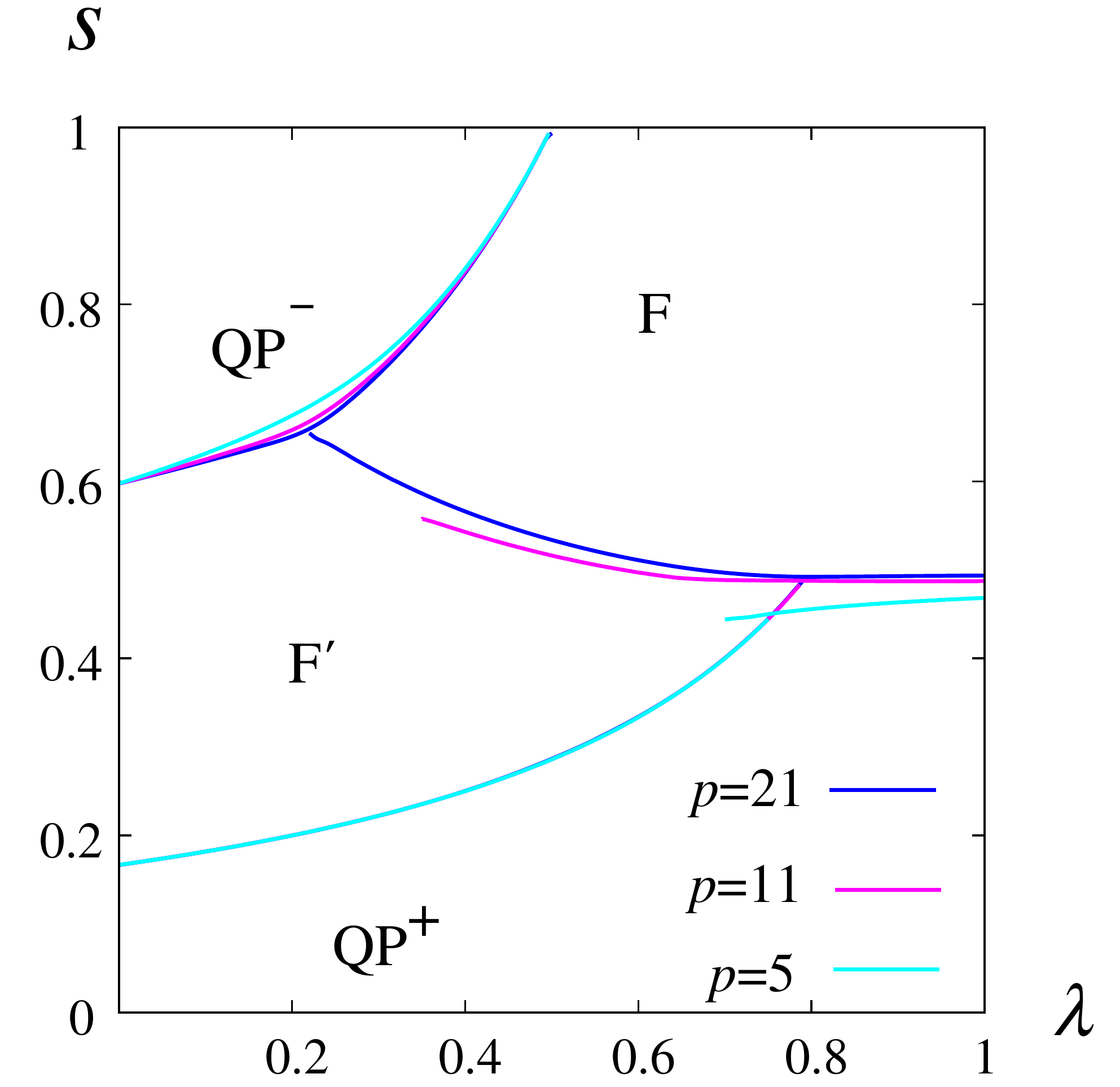}
\end{center}
\caption{$\lambda$-$s$ phase diagram of the non-stoquastic Hamiltonian with $k=5$ (antiferromagnetic $XXXXX$ interactions). The boundary between QP$^+$ (quantum paramagnetic) and F$'$ (ferromagnetic) phases is for second-order phase transitions and all other boundaries represent first-order transitions. A new phase QP$^-$ exists for $k$ odd, where spins point to the $-x$ direction whereas they point to the $+x$ direction in the QP$^+$ phase.  The two ferromagnetic phases F and F$'$ are clearly separated for $p=21$ but not for smaller $p$.}
\label{fig:nonstoquastic_phase_diagram_k5}
\end{figure}

Comparison of Figs. \ref{fig:QA_nonstoquastic_p=5a} and \ref{fig:QA_nonstoquastic_p=5b} suggests that the antiferromagnetic multiple-$X$ term in the Hamiltonian with a large amplitude ($\lambda$ close to 0) would change a first-order phase transition (for large $\lambda$) to second order (small $\lambda$), thus reducing the computation time drastically from exponential to polynomial as a function of the system size. Exhaustive studies have been carried out along this line \cite{Seki2012,Seoane2012}. The results are positive for $p\ge 4$.  Figures \ref{fig:nonstoquastic_phase_diagram_k2} and \ref{fig:nonstoquastic_phase_diagram_k5} show typical examples of the $\lambda$-$s$ phase diagrams.
Figure \ref{fig:nonstoquastic_phase_diagram_k2} is for $k=2$, {\em i.e.} with $XX$ interactions.  For a fixed $p$, the first-order transition at $\lambda=1$ (stoquastic Hamiltonian) extends down to a smaller value of $\lambda$ around the middle of the phase diagram, and then is replaced by a line of second order transitions that continues to $\lambda =0$ and $s=0.33$. The first order transition persists even after the second-order transition line branches out, and there exists a line of first-order transitions within the ferromagnetic phase. We have denoted those two ferromagnetic phases as F and F$'$, though there is no qualitative difference between F and F$'$. In this way, it has been established that a first-order phase transition in the stoquastic Hamiltonian at $\lambda =1$ has been reduced to second order by the effects of the antiferromagnetic multiple-$X$ term with a relatively large amplitude, $\lambda$ close to 0. Quantum annealing starts at $s=0$ and $\lambda$ arbitrary (anywhere on the line $s=0$ at the bottom of the phase diagram) and ends at $s=\lambda=1$ (the upper right corner of the phase diagram), and it is possible to choose a path connecting those points which avoids a first-order transition ({\em i.e.} to go only across the boundary between QP and F$'$ phases). In other words, the  antiferromagnetic multiple-$X$ term, which represents strong quantum effects not to be simulated classically in the standard way, exponentially enhances the efficiency of quantum annealing.

The case of $p=3$ turns out to be an exception in that the first-order transition line persists down to $\lambda=0$. This fact may be interpreted in terms of the Landau theory of phase transitions that there would appear a cubic term in the Landau free energy for the cubic Hamiltonian with $p=3$, which strongly enhances the possibility of first-order transition.

The case of $k=5$ in Fig. \ref{fig:nonstoquastic_phase_diagram_k5} is similar with a few minor differences. There exist two paramagnetic phases, denoted as QP$^+$ and QP$^-$. The former has $\theta =0$  with magnetization vector ${\bm m}=(1,0,0)$ and the latter $\theta=\pi$ with ${\bm m}=(-1,0,0)$. The QP$^-$ phase appears at the top left corner of the phase diagram, where the antiferromagnetic multiple-$X$ term $s(1-\lambda)(m_x)^k$ dominates. The transition is of second order only between QP$^+$ and F$'$ phases.  The first-order transition line within the ferromagnetic phase between F and F$'$ extends toward QP$^-$ at the upper left part of the phase diagram.  For larger $p$, this line reaches the phase boundary of the QP$^-$ phase as seen for $p=21$ in Fig. \ref{fig:nonstoquastic_phase_diagram_k5}. The two ferromagnetic phases, F and F$'$, are then completely separated by a line of first-order transitions.  It is therefore concluded for $k=5$ that a proper choice of annealing path makes it possible to reduce the first-order phase transition for the stoquastic case ($\lambda =1$) to second order (smaller $\lambda$) as long as $p$ is not too large. These features are shared by other values of $k$, even $k$ similar to Fig. \ref{fig:nonstoquastic_phase_diagram_k2} and odd $k$ to Fig. \ref{fig:nonstoquastic_phase_diagram_k5}.

Although the above analyses use only classical variables, it has been shown that quantum statistical-mechanical computations reproduce those phase boundaries quantitatively very faithfully in the thermodynamic limit \cite{Seki2012,Seoane2012}. It has also been confirmed numerically that the energy gap as a function of the system size closes exponentially at first-order phase transitions and polynomially at second order transitions \cite{Seki2012}.  We therefore conclude with confidence that the  antiferromagnetic multiple-$X$ term in the Hamiltonian has the capacity to reduce the computational complexity drastically from exponential to polynomial for the present infinite-range ferromagnetic $p$-spin model.

\subsection{Quantum annealing under the influence of the environment}

It is important to study how the environment affects the behavior of the system.  One of the standard models to describe the interactions of the system with its environment is the following Hamiltonian \cite{Leggett1987}, in which the spin degrees of freedom are coupled linearly with harmonic oscillators representing the environment,
\begin{align}
     H(s,\lambda)=-s\lambda N \Big(\frac{1}{N}\sum_{i=1}^N\sigma_i^z\Big)^p
 &+s(1-\lambda)N\Big(\frac{1}{N}\sum_{i=1}^N\sigma_i^x\Big)^k-(1-s)\sum_{i=1}^N \sigma_i^x \nonumber\\
 &+\frac{1}{\sqrt{N}}\sum_l g_l \big(b_l+b_l^{\dagger}\big)\sum_{i=1}^N \sigma_i^{\alpha} +\sum_l \omega_l b_l^{\dagger}b_l \quad (\alpha =x~{\rm or}~z) .
 \label{coupled-p}
\end{align}
Here $l$ runs over all possible modes of harmonic oscillators, $g_l$ denotes the coupling strength, and $\omega_l$ is the frequency of mode $l$. In this model Hamiltonian (\ref{coupled-p}), the interactions with the environment are assumed to apply uniformly over all sites $i$.  Such a situation may exist when the correlation length of the environment is much larger than the linear size of the system \cite{Breuer2002}.

Following the previous analysis (see also \cite{Sinha2013}), we replace $\sum_i \sigma_i^{\alpha}$ by the classical variable $Nm_{\alpha}$ and rewrite Eq. (\ref{coupled-p}) as
\begin{equation}
H=-s\lambda N (m_z)^p+s(1-\lambda) N(m_x)^k-(1-s)Nm_x+\sum_l \omega_l \tilde{b}_l^{\dagger}\tilde{b}_l - \Lambda N(m_{\alpha})^2,
\label{coupled-p2}
\end{equation}
where
\begin{align}
    \tilde{b}_l^{\dagger}=b_l^{\dagger}+\frac{\sqrt{N}\,g_l}{\omega_l}\, m_{\alpha},\quad
    \tilde{b}_l=b_l+ \frac{\sqrt{N}\,g_l}{\omega_l}\, m_{\alpha},\quad
  \Lambda =\sum_l \frac{g_l^2}{\omega_l}.
  \label{new_variables}
\end{align}
Let us define the spectral density of couplings as
\begin{equation}
J(\omega)=\sum_l g_l^2 \delta (\omega-\omega_l),
\end{equation}
and assume super/normal/sub Ohmic dissipation with cutoff frequency $\omega_{\rm c}$ \cite{Leggett1987}
\begin{equation}
J(\omega)=\alpha \frac{\omega^s}{\omega_{\rm c}^{s-1}}\, e^{-\omega/\omega_{\rm c}}.
\end{equation}
Super, normal, and sub Ohmic cases correspond, respectively, to $s>1$, $s=1$, and $s<1$. We can then write the coefficient $\Lambda$ in Eq. (\ref{new_variables}) as
\begin{equation}
\Lambda =\int _0^\infty \frac{J(\omega)}{\omega}\, d\omega =\alpha \omega_{\rm c}\Gamma (s),
\end{equation}
where $\Gamma (s)$ is the Gamma function.

Equation (\ref{coupled-p2}) reveals that the environment and the spin system are effectively decoupled and can be treated independently. Since we are interested in the ground state, the environment is simply in the vacuum.  The effects of environment to the spin system have been taken into account as the term $-\Lambda N(m_{\alpha})^2$. If we consider for simplicity the case of $k=2$, the environment coupled with the $x$ component of the system $(\alpha=x)$ effectively reduces the coefficient of the  antiferromagnetic multiple-$X$ term from $s(1-\lambda)N$ to $s(1-\lambda)N-\Lambda N$.  This is detrimental to the performance of quantum annealing for the reason discussed in the previous section.  On the other hand, if $\alpha=z$, the normalization of the vector $m_x^2+m_z^2=1$ implies that the final coupling term in Eq. (\ref{coupled-p2}) has a positive contribution to the antiferromagnetic multiple-$X$ term.  Therefore the two types of couplings with the environment $(\alpha =x$ or $z$) give completely the opposite contributions.  More detailed analyses will be given in a forthcoming paper.

\section{Hopfield model}

One may wonder if the above results for the $p$-spin model would apply to more difficult problems.  To answer this question, we have studied the Hopfield model \cite{Seki2015}, which has randomness in interactions and the ground state is non-trivial \cite{Amit1985a,Amit1985b,Amit1987,Nishimori1996}. In the present section, we compare quantum annealing strategies with and without an antiferromagnetic multiple-$X$ term in the Hamiltonian, {\em i.e.} non-stoquastic and stoquastic Hamiltonians, for the Hopfield model.

\subsection{Finite patterns embedded}

The Hopfield model with $p$-body interactions has the Hamiltonian
\begin{equation}
    H_0=-\sum_{i_1 ,\dots ,i_p =1}^N J_{i_1\cdots i_p}
    \sigma_{i_1}^z\cdots \sigma_{i_p}^z,
\end{equation}
where
\begin{equation}
    J_{i_1\cdots i_p}=\frac{1}{N^{p-1}}\sum_{\mu=1}^r
    \xi_{i_1}^{\mu}\cdots \xi_{i_p}^{\mu}
\end{equation}
with each $\xi_{i}^{\mu}$ (representing the state of the $i$th site for the $\mu$th embedded pattern) being $\pm 1$ randomly with equal probability.  The total non-stoquastic Hamiltonian has the same form as before,
\begin{equation}
    H(s,\lambda)=s\lambda H_0 +s(1-\lambda)N
    \Big(\frac{1}{N}\sum_{i=1}^N \sigma_i^x\Big)^2 -(1-s)\sum_{i=1}^N \sigma_i^x.
\end{equation}
The antiferromagnetic multiple-$X$ term has been chosen to be quadratic ($k=2$ in the notation of the previous section) for simplicity. We first discuss the case with the number of embedded patterns $r$ finite.  

It is impossible to apply the simple classical method used in the $p$-spin ferromagnetic model because of the complexity of interactions.  Quantum statistical-mechanical techniques have been exploited in \cite{Seki2015}, by which the quantum system is reduced to a corresponding classical Ising model by the Suzuki-Trotter decomposition.  We refer the reader to \cite{Seki2015} for details and write the resulting energy per site as a function of the order parameters,
\begin{align}
e(m_1^z,\cdots,m_r^z,m^x)&=(p-1)s\lambda \sum_{\mu =1}^r \big(m_{\mu}^z\big)^p - s(1-\lambda)(m^x)^2\nonumber\\
&-\left[ \sqrt{\big(ps\lambda \sum_{\mu} (m_{\mu}^z)^{p-1}\xi^{\mu}\big)^2 +(1-s-2s(1-\lambda)m^x)^2} \right],
\label{e_Hopfield1}
\end{align}
where the square brackets stand for the average over the random variables $\{\xi^{\mu}\}$. Notice that the index $i$ of $\xi_i^{\mu}$ has disappeared in the above equation due to the infinite-range (mean-field) characteristics of the Hopfield model.

The parameter $m^x$ has the same meaning as in the $p$-spin ferromagnetic model, the $x$ component of the averaged spin operator. The other parameter $m_{\mu}^z$ represents the overlap (or similarity) of the $z$ component of the Pauli matrix with the $\mu$th embedded pattern,
\begin{equation}
m_{\mu}^z=\frac{1}{N}\left[\sum_{i=1}^N \xi_i^{\mu}\langle \sigma_i^z\rangle \right],
\end{equation}
where the angular brackes $\langle \cdots \rangle$ denote the average by the ground-state wave function.

The energy of Eq. (\ref{e_Hopfield1}) is to be minimized with respect to the order parameters $m^x$ and $m_{\mu}^z$.  There exist a large number of candidate states that are the solutions to the self-consistent equation obtained as the vanishing condition of the derivatives of the energy with respect to the order parameters. It is known in the classical Hopfield model ($s=\lambda=1$) at finite temperatures that the simplest non-trivial solution, $m_1^z>0$ and $m_2^z=\cdots =m_{r}^z=0$, has the lowest free energy and is realized at low temperatures, in addition to the paramagnetic solution (all $m_{\mu}^z=0$) valid at high temperatures \cite{Amit1985a}. Almost the same turns out to be the case in the quantum Hopfield model \cite{Nishimori1996,Seki2015}, the trivial differences being that the energy, not the free energy, is to be minimized and that the quantum paramagnetic state has $m^x>0$.  

When only $m_1^z$ is finite with all other $m_{\mu}^z$'s being zero, the energy Eq. (\ref{e_Hopfield1}) turns out to have exactly the same form as the corresponding energy of the $p$-spin ferromagnetic model analyzed by the quantum statistical-mechanical methods. Thus the analyses of the previous section apply directly. This is the same situation as in the classical Hopfield model \cite{Amit1985a}. We may then conclude that the  antiferromagnetic multiple-$X$ term helps the Hopfield model avoid first-order phase transitions in the process of quantum annealing exactly in the same way as in the $p$-spin ferromagnetic system. It has thus been established that the antiferromagnetic multiple-$X$ term exponentially improves the efficiency of quantum annealing even in the presence of randomness in interactions.

\subsection{Many numbers embedded (I)}

When $r$, the number of patterns embedded, increases with the system size $N$, the situation becomes dependent on $p$. We discuss the case of $p=2$ in this section.

When $p=2$ and $r$ is supposed to increase with $N$, $r$ turns out to be proportional to $N$, $r=\alpha N$, in order for the free energy to be extensive. Under this condition, the free energy for arbitrary temperature can be evaluated using the standard techniques from quantum statistical mechanics, the Suzuki-Trotter decomposition, the replica method under replica-symmetric ansatz, and the static approximation to drop the Trotter-number dependence of order parameters \cite{Nishimori1996,Seki2015}. The resulting free energy per site in the zero-temperature limit (the ground state energy) is
\begin{align}
    e(m, q, m^x)&=\frac{1}{2}\, s\lambda m^2-s(1-\lambda)(m^x)^2-\frac{\alpha}{2}\, s\lambda+\frac{\alpha}{2}\, \tilde{q}C\nonumber\\
    &-\int Dz\,\sqrt{(s\lambda m+\sqrt{\alpha\tilde{q}})^2+(1-s-2s(1-\lambda)m^x)^2},
\end{align}
where
\begin{align}
    \tilde{q}&=\frac{s^2 \lambda^2 q}{(1-s\lambda C)^2}\\
    C&=\int Dz\frac{(1-s-2s(1-\lambda)m^x)^2}{\big((s\lambda m+\sqrt{\alpha \tilde{q}}\, z)^2+(1-s-2s(1-\lambda)m^x)^2\big)^{3/2}}
\end{align}
with $Dz=\exp (-z^2/2)\, dz/\sqrt{2\pi}$.  The parameters $m, q, m^x$ denote the overlap, the spin glass order parameter, and the magnetization along the $x$ axis, 
\begin{equation}
m=\left[ \xi_i^1 \langle \sigma_i^z\rangle \right],~
q= \left[ \langle \sigma_i^z\rangle^2\right],~
m^x= \left[\langle \sigma_i^x \rangle \right].
\end{equation}
It has been assumed that the solution with only one of the embedded patterns being retrieved is more stable than other possibilities, as in the case of the classical Hopfield model \cite{Amit1985a,Amit1985b,Amit1987}.

\begin{figure}[h!]
\begin{center}
\includegraphics[width=.4\textwidth]{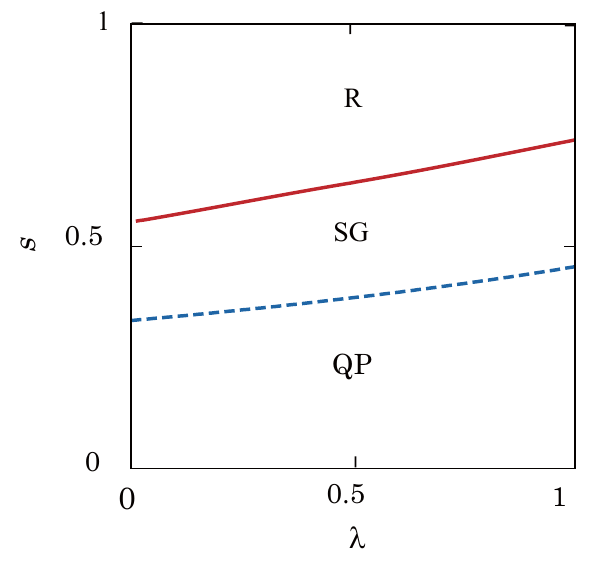}
\end{center}
\caption{Phase diagram of the Hopfield model with $p=2$ and $r=0.04N$. The boundary between the quantum paramagnetic phase (QP) and the spin glass phase (SG) is for second order transition and the boundary between the SG and the retrieval phase (R) is of first order.}\label{fig:Hopfield_phase_diagram2}
\end{figure}

Extremization conditions of $e$ with respect to $m, q$ and $m^x$ and comparison of energy values among different solutions lead to the phase diagram of Fig. \ref{fig:Hopfield_phase_diagram2}.
There exist three phases, quantum paramagnetic (QP) ($m=q=0$), spin glass (SG) ($m=0, q>0$), and retrieval (R) ($m>0, q>0$).  In the stoquastic case with $\lambda =1$, it is known that the phase transition between the spin glass and retrieval phases is of first order \cite{Nishimori1996}.  As shown in Fig. \ref{fig:Hopfield_phase_diagram2}, this first-order transition persists even when an antiferromagnetic multiple-$X$ term is introduced down to $\lambda=0$. It is impossible to reach the final state $s=\lambda=1$ through a path that avoids a first-order transition starting from the initial state with $s=0$.  The spin-glass phase covering the middle of the phase diagram causes an essential difficulty in the present case \footnote{It has been pointed out in \cite{Knysh2016}  that a different type of difficulty exists within the spin glass phase of the Hopfield model when the random variables $\xi_i^{\mu}$ are Gaussian-distributed, not binary as in the present paper.}.

\subsection{Many numbers embedded (II)}
\begin{figure}[h!]
\begin{center}
\includegraphics[width=.37\textwidth]{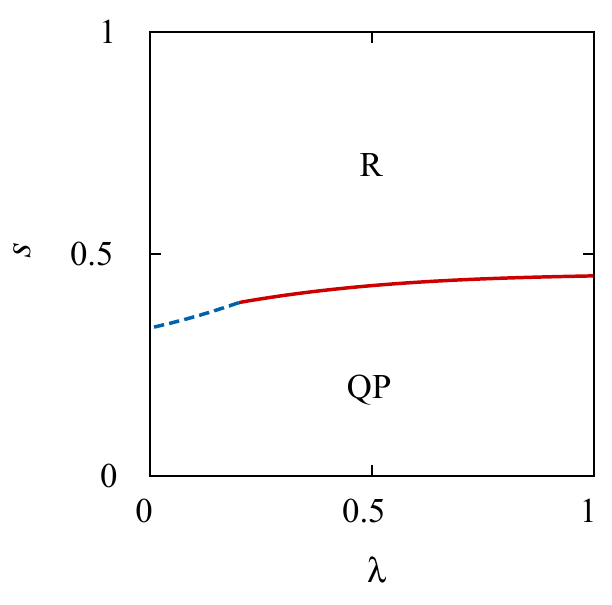}
\end{center}
\caption{Phase diagram of the Hopfield model with $p=4$ and $r=0.04N^3$. The first-order transition in red is replaced by second-order transitions in blue for small $\lambda$.}\label{fig:Hopfield_phase_diagram3}
\end{figure}

We next discuss the case of $p\ge 3$.  Again, the standard quantum statistical-mechanical method can be applied to the analysis of the model with $p\ge 3$ \cite{Seki2015}. Since the computations are straightforward but quite lengthy, we refer the reader to \cite{Seki2015} for details \footnote{Notice that the replica symmetric ansatz \cite{nishimori_book_sg} is used in the calculations. Our experience in the simple quantum Hopfield model in a transverse field suggests that the replica symmetry breaking takes place only in a very limited region in the phase diagram \cite{Nishimori1996}, and we expect it to be reasonable to assume a similar situation in the present case as well.}. The result for the ground-state energy as a function of order parameters is,
\begin{align}
    e(m,q,m^x)&=s\lambda (p-1) m^p -s(1-\lambda)(m^x)^2 +\frac{\alpha}{2}p (p-1) (s\lambda)^2 C q^{p-1} \nonumber\\
    &-\int Dz\, \sqrt{\big(s\lambda (p m^{p-1}+\sqrt{\alpha p q^{p-1}}\, z)\big)^2+\big(1-s-2s(1-\lambda)m^x\big)^2},
\end{align}
where
\begin{equation}
C=\int Dz\, \frac{(1-s-2s(1-\lambda)m^x)^2}{\Big\{\big(s\lambda (pm^{p-1}+\sqrt{\alpha p q^{p-1}}\, z)\big)^2 +(1-s-2s(1-\lambda)m^x))^2\Big\}^{3/2}}.
\end{equation}
The extremization condition of the energy leads to a set of self-consistent equations for the order parameters, the solutions to which indicate possible phases at each point in the $\lambda$-$s$ phase diagram. It turns out that the spin glass phase always has a higher energy than other phases, the retrieval phase and the paramagnetic phase, and is not realized as a stable phase for $p\ge 3$.  The transition between the retrieval and paramagnetic phases for the stoquastic model ($\lambda =1$) is of first order.  The introduction of the antiferromagnetic multiple-$X$ term ($\lambda <1$) leads to replacement of this first-order transition by a second-order transition below a threshold value of $\lambda$  provided that $p>3$.  An example of $p=4$ is depicted in Fig. \ref{fig:Hopfield_phase_diagram3}.
For $p=3$, the first-order transition continues to exist up to $\lambda =0$ as was the case without randomness.

We have therefore established that the  antiferromagnetic multiple-$X$ term has the effect of drastically enhancing the computational efficiency even in some cases with randomness.

\section{Conclusion}

We have shown that antiferromagnetic multiple-$X$ terms reduce a first-order quantum phase transition to second order in the infinite-range ferromagnetic $p$-spin model as well as in the quantum Hopfield model.  This means that the efficiency of quantum annealing in its formulation as adiabatic quantum computation is exponentially enhanced by the antiferromagnetic multiple-$X$ term, which renders the Hamiltonian non-stoquastic. Although not shown explicitly in the present article, it has been confirmed numerically for the ferromagnetic $p$-spin model in \cite{Seki2012} that the minimum gap at the phase transition indeed closes exponentially or polynomially according to the order of phase transition.  It is reasonable to expect that the same holds for the Hopfield model.  Since a non-stoquastic Hamiltonian cannot be simulated efficiently on classical computers in the standard quantum Monte Carlo simulation, it may be interpreted to represent strong quantum effects.  We may therefore conclude that the exponential enhancement of the efficiency for quantum annealing is achieved in the present models by strong quantum effects. These are the first cases, as far as the authors are aware of, where such a conclusion has been drawn by analytical methods. Notice in this relation that numerical evidence of related nature was presented in \cite{Farhi2002,Crosson2014,Hormozi2016}.  Our conclusion does not necessarily exclude the existence of other efficient numerical methods to study a given non-stoquastic Hamiltonian including, possibly, the spin-vector dynamics \cite{Smolin2014,Albash2015,Owerre2014,Muthukrishnan2016} or even the simple steepest descent method.

We have also shown for the ferromagnetic $p$-spin model that certain types of system-environment couplings either enhance or reduce the effect of the antiferromagnetic multiple-$X$ term depending on the component of spin operators appearing in the coupling term.  The argument leading to this conclusion crucially depends on the special property of the ferromagnetic $p$-spin model that the spin Hamiltonian commutes with the total spin operator. It is an interesting question whether or not similar behavior can be observed in other cases.

It should be remembered that antiferromagnetic multiple-$X$ terms discussed in the present paper are far from versatile to enhance the efficiency. Indeed, the first-order transitions in the ferromagnet with $p=3$ and the Hopfield model with $p=2$ have been shown not to be relaxed to second order.  It has also been known that the first-order transition in the $p$-body interacting random-field Ising model persists in the presence of antiferromagnetic multiple-$X$ terms if the distribution of random field is bimodal \cite{Ichikawa2014}.

One may wonder if there is any other way to show that an antiferromagnetic multiple-$X$ term indeed represents strong quantum effects by a more direct method, not via the impossibility of the standard quantum Monte Carlo technique. We are now investigating this problem, and the result will soon be published \cite{Susa2016}.  One of the hints may lie in the sign of coefficients of the ground-state wave function in the standard computational basis. For a stoquastic Hamiltonian, the coefficients can be chosen to be all non-negative, according to the Perron-Frobenius theorem. This leads to the natural interpretation of the (squared) magnitude of a coefficient as the probability. If, in contrast, the Hamiltonian is non-stoquastic, some of the coefficients can be negative or even complex, and the conventional interpretation of the squared absolute value of the coefficient as the probability does not necessarily fit very well to our (classical) intuition.  Whether or not this fact suggests {\em strong} quantum effects needs further scrutiny.

It is an interesting question how far the present results for the infinite-range fully-connected models apply to more realistic problems with relatively sparse connections, e.g. a problem on a finite-dimensional lattice with short-range interactions.  It is of course difficult to say something with confidence without explicit evidence. Nevertheless, our experience in the physics of phase transitions suggests that a mean-field analysis often provides reliable results also for finite-dimensional systems with a finite number of connections per site as far as qualitative descriptions are concerned \cite{Nishimori_book}.  It would come as a surprise if this general rule of thumb does not apply to the present case.  It is worth an effort to investigate if and how  antiferromagnetic multiple-$X$ terms widen the energy gap in systems with finite connections.  We are studying a related problem, and the results will be published before too long \cite{Okuyama2016}.

\begin{acknowledgments}
Most of the technical results described in this article appeared in Refs. \cite{Seki2012,Seki2015,Seoane2012} albeit from a little different viewpoint than presented here. One of the authors (H.N.) sincerely thanks Yuya Seki and Beatriz Seoane for stimulating collaboration. Also acknowledged are useful comments by Tameem Albash, Jacob Biamonte, Eddie Farhi, Itay Hen, Layla Hormozi, Helmut Katzgraber, and Daniel Lidar. This work was funded by the ImPACT Program of Council for Science, Technology and Innovation, Cabinet Office, Government of Japan, and by the JPSJ KAKENHI Grant No. 26287086. 
\end{acknowledgments}

\bibliography{refs}

\end{document}